\documentclass[%
 reprint,
preprintnumbers,
nofootinbib,
 amsmath,amssymb,
 aps,
]{revtex4-1}
\usepackage{slashed}
\usepackage{graphicx}
\usepackage{dcolumn}
\usepackage{bm}

\usepackage{amsfonts}
\usepackage{latexsym}
\usepackage{dsfont}
\usepackage{amsmath}
\usepackage{amssymb}
\usepackage{amssymb}
\usepackage{amsthm}
\usepackage{setspace}

\allowdisplaybreaks

\def\dd{\text{d}}
\def\lp{\lambda_+}
\def\lm{\lambda_-}

\begin{document}

\preprint{DMUS--MP--22/08}

\title{Extending the non-relativistic string AdS coset}
\author{Andrea Fontanella}
\email[E-mail: ]{\tt a.fontanella.physics@gmail.com} 
\affiliation{\vspace{2mm} Perimeter Institute for Theoretical Physics, \\
Waterloo, Ontario, N2L 2Y5, Canada}

\author{Juan Miguel Nieto Garc\'ia}
 \email{E-mail: \tt j.nietogarcia@surrey.ac.uk}
\affiliation{\vspace{2mm} Department of Mathematics, University of Surrey,
\\ Guildford, GU2 7XH, UK}



\begin{abstract}
Inspired by Lie algebra expansion, we consider an extension of the algebra introduced in \href{https://arxiv.org/abs/2203.07386}{arXiv:2203.07386} for the non-relativistic string coset action in AdS$_5\times$S$^5$. We show that the extended algebra admits a non-degenerate inner product which is adjoint invariant under the full extended algebra. Furthermore, we provide a finite-dimensional representation of the extended algebra.   
 
\end{abstract}

\maketitle

Non-relativistic (NR) strings may be regarded potentially interesting from the holographic point of view, since they propagate in string Newton-Cartan geometries, which are non-Lorentzian. Therefore they represent a novel arena where to study non-AdS holography, which is currently still poorly understood. NR strings have first been introduced in flat space \cite{Gomis:2000bd, Danielsson:2000gi}, and later also in AdS$_5\times$S$^5$ \cite{Gomis:2005pg}.
NR strings are Weyl anomaly free in ten dimensional spacetimes \cite{Gomis:2019zyu}, and setting to zero the beta function produces NR gravity equations of motion for the background geometry. For a recent review of NR strings, see e.g. \cite{Oling:2022fft}. Few topics of NR strings in AdS$_5\times$S$^5$ have been studied so far, such as classical solutions \cite{Fontanella:2021btt}, semiclassical expansion in light-cone gauge \cite{Fontanella:2021hcb} and the coset formulation of the action and its Lax representation \cite{Fontanella:2022fjd}.

In this letter, we focus on further understanding the coset action for NR strings in AdS proposed in \cite{Fontanella:2022fjd}. We show that the particular Lie algebra expansion considered in \cite{Fontanella:2020eje} naturally suggests considering non-central extensions, denoted by $Z_{ab},  Z_{a'b'}$, in the coset construction. Although the fields introduced by such extensions are unphysical, since they appear both in the coset numerator and denominator, the extended algebra admits a non-degenerate inner product which is adjoint invariant under the full algebra. It is worthwhile noticing that an inner product with such nice properties cannot be constructed for the original algebra without the extensions $Z_{ab},  Z_{a'b'}$. Furthermore, inspired by Lie algebra expansion, we are able to provide a representation for the generators of the extended algebra.

\section{The coset action of \cite{Fontanella:2022fjd}}
The action for NR strings in AdS$_5\times$S$^5$, in the set of coordinates originally used in \cite{Gomis:2005pg}, is 

\null
{\small
\null \vspace{-12mm} 
\begin{equation}\label{eq:NRAdS}
\begin{aligned}
S &= - \frac{T}{2} \int \dd^2 \sigma \, \bigg[\gamma^{\alpha\beta}\bigg( x^a x_a (-\partial_{\alpha} x^0 \partial_{\beta} x^0 + \cos^2 x^0 \partial_{\alpha} x^1 \partial_{\beta} x^1) \\
&+\partial_{\alpha} x^a \partial_{\beta} x_a + \partial_{\alpha} x^{a'} \partial_{\beta} x_{a'} \bigg)+ \varepsilon^{\alpha\beta} \bigg( (\lp e_{\alpha}{}^+ + \lm e_{\alpha}{}^-) \partial_{\beta} x^0 \\
&+ (\lp e_{\alpha}{}^+ - \lm e_{\alpha}{}^-) \cos x^0 \partial_{\beta} x^1 \bigg) \bigg] \ ,
\end{aligned}
\end{equation}}
where $T$ is the string tension, $\sigma^{\alpha} = (\tau, \sigma)$, with $\alpha = 0, 1$, are the string world-sheet coordinates, $\gamma^{\alpha\beta} \equiv \sqrt{-h} h^{\alpha\beta}$ is the Weyl invariant combination of the inverse world-sheet metric $h^{\alpha\beta}$ and $h =$ det$(h_{\alpha\beta})$, and $e_{\alpha}{}^{\pm}$ are (the light-cone components of) the worldsheet zweibein. $x^A$, with $A=0,1$, are longitudinal coordinates originating from AdS$_5$, while $x^a$ and $x^{a'}$, with $a, b, ... = 2, 3, 4$ and $a', b', ... = 1, ..., 5$, are transverse coordinates originating from AdS$_5$ and S$^5$ respectively, which are contracted with $\delta_{ab}$, $\delta_{a'b'}$. $\lambda_{\pm}$ are non-dynamical scalar Lagrange multiplier fields.

The idea now is to rewrite the degrees of freedom of this action in terms of a coset sigma model.

 Typically, the coset numerator $G$ is taken to be the isometry group of the target space. Such group is not yet fully determined, although known to be infinite dimensional \cite{Bagchi:2009my}. In the flat space case, the isometry group is also infinite dimensional but, on the contrary of the AdS case, it is fully known \cite{Batlle:2016iel}. Since it is unpractical to work with infinite dimensional algebras, it is convenient to restrict to a consistent finite truncation. In the flat space, a particular extension of the string Bargmann algebra can be identified as the finite subalgebra we are interested in\footnote{There is at least another finite subalgebra, known as string Newton-Cartan algebra, which one may consider in order to construct a NR string coset action in flat space. However, such choice gives equations of motion less suitable for a Lax representation.}, as it contains (a subalgebra of) the local symmetry expected on a NR string action. This subalgebra can also be generated by a particular technique known as Lie algebra expansion. This same technique may then be applied to the AdS case, and fix the coset numerator without requiring any particular knowledge of the infinite dimensional isometry algebra. 
The coset denominator $H$ is then fixed accordingly by a counting of degrees of freedom. 

Concretely, the coset numerator algebra $\mathfrak{g}$ considered in the AdS case of \cite{Fontanella:2022fjd} is the direct sum of a particular extension of a string Newton-Hooke and an Euclidean algebra. 
The extended string Newton-Hooke algebra is spanned by a longitudinal boost $M$, longitudinal translations $H_A$, transverse rotations $J_{ab}$, transverse translations $P_a$, string-Galilei boosts $G_{Ab}$, and non-central extensions $Z_A$ and $Z$. These generators have the following non-vanishing commutation relations  
\begin{equation} \label{ext_NH_algebra}
	\begin{aligned}
	{}[J_{ab}, J_{cd}] & = \delta_{bc} J_{ad} - \delta_{ac} J_{bd} + \delta_{ad} J_{bc} - \delta_{bd} J_{ac} \,, \\
	[J_{ab},P_{c}] & =  \delta_{bc} P_{a} - \delta_{ac} P_{b}\,, \\
    [J_{ab},G_{Ac}] & = \delta_{bc} G_{Aa} - \delta_{ac} G_{Ab} \,,\\
	[G_{Aa}, G_{Bb}] & = \delta_{ab} \varepsilon_{AB} Z\,, \\ 
	[M,G_{Aa}]  & = -\varepsilon_A{}^B G_{Ba}\,, \\
	[G_{Aa}, P_{b}] & = \delta_{ab} Z_A\,,\\
	[G_{Aa},H_B]  & = -\eta^{}_{AB} P_{a}\,, \\ 
    [H_A,Z_B] & = -\varepsilon_{AB} Z \,, \\
    [M,Z_A] & = -\varepsilon_A{}^B Z_B\,, \\ 
	[Z,H_A]  & = -\varepsilon_A{}^B Z_B\,,\\
	[M,H_A] & = -\varepsilon_A{}^B H_B\,, \\
	[H_A, H_B]& = -\varepsilon_{AB} M \, ,\\
	[H_A, P_b] &= G_{Ab} \ . 
	\end{aligned}
\end{equation}
The Euclidean algebra is spanned by spatial translations $P_{a'}$ and spatial rotations $J_{a'b'}$, with commutation relations
\begin{equation}
\label{Euclidean}
\begin{aligned}
{}[P_{a'}, J_{b'c'}] &= \delta_{a'b'} P_{c'} - \delta_{a'c'} P_{b'} \, , \\
[J_{a'b'}, J_{c'd'}] &= \delta_{b'c'} J_{a'd'} - \delta_{a'c'} J_{b'd'} + \delta_{a'd'} J_{b'c'} - \delta_{b'd'} J_{a'c'} \, .
\end{aligned}
\end{equation}
The coset denominator algebra $\mathfrak{h}$ is taken to be everything that generates the numerator except for $H_A, P_a, P_{a'}$.

As showed in \cite{Fontanella:2022fjd}, the action (\ref{eq:NRAdS}) may be rewritten in a coordinate-free language in terms of a Maurer Cartan (MC) 1-form $A = g^{-1} \dd g$ with $g \in G$, an external current $\Lambda_{\alpha} = \lm e_{\alpha}{}^-  \, Z_+ + \lp e_{\alpha}{}^+  \, Z_-$  which encodes the $\lambda_{\pm}$ fields, and a symmetric bilinear inner product $\langle \cdot, \cdot \rangle$ on $\mathfrak{g}$, 
\begin{equation}
\label{eq:NR_coset_action}
S^{G/H} = - \frac{T}{2} \int \dd^2 \sigma \, \gamma^{\alpha\beta}  \langle J^{(1)}_{\alpha}, J^{(1)}_{\beta} \rangle \ ,
\end{equation}
where $J_{\alpha}$ is defined as $J_{\alpha} = A_{\alpha} + \gamma_{\alpha\beta} \varepsilon^{\beta\gamma} \Lambda_{\gamma}$, and $J^{(1)}$ is the projection of $J$ into the grading 1 subspace with respect to the $\mathbb{Z}_2$ automorphism, which induces the split
\begin{equation}
\mathfrak{g} = \mathfrak{g}^{(0)} \oplus \mathfrak{g}^{(1)} \, . 
\end{equation}
The inner product is demanded to be adjoint invariant under $\mathfrak{g}^{(0)} \equiv \mathfrak{h}\setminus \{Z_A\}$. The most generic symmetric bilinear with such property is
	\begin{equation}\label{inner_product_AdS}
	\begin{aligned}
	\langle P_a, P_b \rangle &=  \omega_1  \delta_{ab} \ , &
	\langle H_{\pm}, Z_{\mp} \rangle &=  -\omega_1/2 \ , \\
	\langle H_+, H_- \rangle &= \omega_2 \ , &
	\langle M, M \rangle &= \omega_3 \ , \\
	\langle J_{ab}, J_{cd} \rangle &= \omega_4 \delta_{[a[c} \delta_{b]d]}  \ , &
	\langle P_{a'}, P_{b'} \rangle &=  \omega_6  \delta_{a'b'} \ , \\
	\langle J_{a'b'}, J_{c'd'} \rangle &= \omega_5 \delta_{[a'[c'} \delta_{b']d']}  \ . &
	\end{aligned}
\end{equation}
In order for (\ref{eq:NR_coset_action}) to reproduce precisely (\ref{eq:NRAdS}), one needs to set\footnote{This condition is coordinate-free, and required by the expected string Newton-Cartan structure of a NR string action.} $\omega_2 =0$ and choose the set of coordinates 
\begin{equation}
\label{GGK_coset}
g = g_{l} g_{t}, \qquad g_l = e^{x^1 H_1} e^{x^0 H_0}, \qquad  g_t = e^{x^a P_a + x^{b'} P_{b'}} \ . 
\end{equation} 
Although (\ref{inner_product_AdS}) is degenerate in the full algebra $\mathfrak{g}$, it is non-degenerate in the generators carrying physical d.o.f., i.e. in $\mathfrak{g}\setminus \mathfrak{h}$. It is worthwhile noticing that enforcing adjoint invariance under a larger algebra than $\mathfrak{g}^{(0)}$ makes the inner product degenerate also in the $\mathfrak{g}\setminus \mathfrak{h}$ subspace.

The equations of motion coming from the action (\ref{eq:NR_coset_action}) admit a Lax pair, 
\begin{equation}
\label{eq:lax}
\mathcal{L}_\alpha = \ell_0 A_\alpha^{(0)} + \ell_1 A_\alpha^{(1)} + \ell_2 \gamma_{\alpha\beta}\varepsilon^{\beta \gamma} J_\gamma^{(1)} \, , 
\end{equation}
where the parameters need to satisfy the relation $\ell_1^2 - \ell_2^2 = 1$. The Lax representation of the equations of motion holds on the constraint surface defined by the equations of motion for $\lambda_{\pm}$.

\section{Extending the algebra}

Although the algebra given in (\ref{ext_NH_algebra}) and (\ref{Euclidean}) is obtained from Lie algebra expansion of $\mathfrak{so} (2,4) \oplus \mathfrak{so}(6)$, it is not the most generic algebra that can be obtained from this procedure. A more general Lie algebra expansion was given in \cite{Fontanella:2020eje}, schematically 
\begin{equation}\label{LAE}
\begin{aligned}
J_{AB} &\rightarrow \varepsilon_{AB} ( M + \epsilon^2 Z ) \, ,\\
J_{Aa} &\rightarrow \epsilon G_{Aa}\, ,\\
J_{ab} &\rightarrow J_{ab} + \epsilon^2 Z_{ab} \, , \\
P_{A}&\rightarrow H_A + \epsilon^2 Z_A \, , \\
P_{a} &\rightarrow \epsilon P_a \, ,\\
J_{a'b'}&\rightarrow J_{a'b'} + \epsilon^2 Z_{a'b'}\, , \\
P_{a'}&\rightarrow \epsilon P_{a'} \, , 
\end{aligned}
\end{equation}
where on the l.h.s. are the generators of $\mathfrak{so} (2,4) \oplus \mathfrak{so}(6)$, and $\epsilon$ is the expansion parameter. The commutation relations for the expanded algebra are inherited from the commutation relations of the parental algebra via power matching in $\epsilon$, together with consistency conditions of the flatness of MC 1-forms.

The expansion (\ref{LAE}) generates an algebra, $\hat{\mathfrak{g}}$, which contains the same set of generators as the one considered in the previous section, supplemented with $\{Z_{ab}, Z_{a'b'}\}$. The non-vanishing commutation relations of $\hat{\mathfrak{g}}$ are given by (\ref{ext_NH_algebra}) and (\ref{Euclidean}) after replacing the old $[P_a, P_b]$, $[G_{Aa}, G_{Bb}]$, $[P_{a'}, P_{b'}]$ by
\begin{equation}
\begin{aligned}
\null [P_a, P_b] &= Z_{ab} \, ,\\
	[G_{Aa}, G_{Bb}] & = \delta_{ab} \varepsilon_{AB} Z - \eta_{AB} Z_{ab} \,, \\ 
	[P_{a'}, P_{b'}] &= - Z_{a'b'} \, ,
\end{aligned}
\end{equation}
and supplementing them with
\begin{equation}
\begin{aligned}
\null [J_{ab}, Z_{cd}] & = \delta_{bc} Z_{ad} - \delta_{ac} Z_{bd} + \delta_{ad} Z_{bc} - \delta_{bd}
 Z_{ac} \,, \\
 [J_{a'b'}, Z_{c'd'}] &= \delta_{b'c'} Z_{a'd'} - \delta_{a'c'} Z_{b'd'} + \delta_{a'd'} Z_{b'c'} - \delta_{b'd'} Z_{a'c'} \, .
\end{aligned}
\end{equation}

For the new algebra $\hat{\mathfrak{g}}$, we construct the most generic bilinear inner product which is adjoint invariant under the full $\hat{\mathfrak{g}}$, 
	\begin{equation}\label{inner_product_extended_algebra}
	\begin{aligned}
	\langle M, Z \rangle &= \phi_1\ , &
	\langle H_{\pm}, Z_{\mp} \rangle &=  -\phi_1/2 \ , \\
	\langle H_+, H_- \rangle &= - \phi_2 /2 \ , &
	\langle M, M \rangle &= \phi_2 \ , \\
	\langle J_{ab}, Z_{cd}  \rangle &= -2 \delta_{[a[c} \delta_{b]d]}  \phi_1  \ , &
	\langle G_{\pm a}, G_{\mp b} \rangle &=   \delta_{ab}  \phi_1/2 \ , \\
	\langle J_{ab}, J_{cd} \rangle &=  \delta_{[a[c} \delta_{b]d]}  \phi_3 \ , &
	\langle P_a, P_b \rangle &=    \delta_{ab} \phi_1 \ , \\
	\langle J_{a'b'}, J_{c'd'} \rangle &=  \delta_{[a'[c'} \delta_{b']d']} \phi_4 \ , &
	\langle P_{a'}, P_{b'} \rangle &=   \delta_{a'b'} \phi_5  \ , \\
	\langle J_{a'b'}, Z_{c'd'} \rangle &= 2 \delta_{[a'[c'} \delta_{b']d']} \phi_5 \ , &
	\end{aligned}
\end{equation}
where $\phi_i$ are arbitrary constants. Although both $\mathfrak{g}$ and $\hat{\mathfrak{g}}$ are not semi-simple, in contrast to what happens to $\mathfrak{g}$, the inner product (\ref{inner_product_extended_algebra}) is: (i) adjoint invariant under the full algebra $\hat{\mathfrak{g}}$, and (ii) non-degenerate in $\hat{\mathfrak{g}}$.

\section{A matrix representation}

Inspired by the Lie algebra expansion (\ref{LAE}), we are able to construct a finite dimensional matrix representation for the generators of $\hat{\mathfrak{g}}$. Each generator has a level given by its power in $\epsilon$ coming from the expansion. In particular, we can split $\hat{\mathfrak{g}}$ into three subspaces $\hat{\mathfrak{g}} = \hat{\mathfrak{g}}_0 \oplus \hat{\mathfrak{g}}_1 \oplus \hat{\mathfrak{g}}_2$,
\begin{equation}
\begin{aligned}
\hat{\mathfrak{g}}_0 &= \{M, H_A, J_{ab}, J_{a'b'} \} \, , \\
\hat{\mathfrak{g}}_1 &= \{G_{Aa}, P_a, P_{a'} \} \, , \\
 \hat{\mathfrak{g}}_2 &= \{Z, Z_A, Z_{ab},  Z_{a'b'} \} \, . 
\end{aligned}
\end{equation}
Note that the commutation relations are graded with respect to this level, 
\begin{equation}\label{LAE_grading}
[\hat{\mathfrak{g}}_n , \hat{\mathfrak{g}}_m] \subset \hat{\mathfrak{g}}_{n+m} \, , 
\end{equation}
where $\hat{\mathfrak{g}}_{n} = \emptyset$ for $n >2$. A representation $\hat{\rho}$ for $\hat{\mathfrak{g}}$ can be constructed by taking a representation for $\mathfrak{so} (2,4) \oplus \mathfrak{so}(6)$, indicated by $\rho$, in tensor product with linear maps $\omega_i$ acting on a 3-dimensional vector space whose basis carry the generator level. These linear maps are required to respect the grading condition (\ref{LAE_grading}) and are chosen to be
\begin{equation}
\omega_0 = \begin{pmatrix}
1 & 0 & 0\\
0 & 1 & 0 \\
0 & 0 & 1
\end{pmatrix},  \  
\omega_1 = \begin{pmatrix}
0 & 0 & 0\\
1 & 0 & 0 \\
0 & 1 & 0
\end{pmatrix} , \ 
\omega_2 = \begin{pmatrix}
0 & 0 & 0\\
0 & 0 & 0 \\
1 & 0 & 0
\end{pmatrix} . \ 
\end{equation}
Concretely, for a given generator $\hat{X} \in \hat{\mathfrak{g}}_i$, its representation is given by   
\begin{equation}
\hat{\rho}(\hat{X}) \equiv \omega_i \otimes \rho ( X )
\end{equation}
where $X\in \mathfrak{so} (2,4) \oplus \mathfrak{so}(6)$ is the parent generator associated with $\hat{X}$ from the expansion (\ref{LAE}).  

If we choose $\rho$ to be the spinorial representation, see e.g. \cite{Arutyunov:2009ga} for our conventions, an inner product on $\hat{\rho}$, invariant under the adjoint action of $\hat{\mathfrak{g}}$, is constructed as follows
\begin{equation}
\langle \hat{\rho}(\hat{X}), \hat{\rho}(\hat{Y}) \rangle \equiv \text{STr} \left[\left( (a \,\omega_0 +  b \, \omega_1 + c\, \omega_2)^t \otimes \mathbf{1}\right) \hat{\rho}(\hat{X}) \hat{\rho}(\hat{Y} )\right] \, , 
\end{equation}
where the prefactors $\omega_i^t \otimes \mathbf{1}$ are responsible for undoing the level shift, and $a, b, c$ are arbitrary constants. The supertrace `STr' is defined as 
\begin{equation}
\text{STr} (m \otimes \mathcal{M}) \equiv \text{Tr}(m) \text{STr}(\mathcal{M}) \ , 
\end{equation}
where the supertrace in the spinorial representation is computed in the usual way, i.e. $\text{STr}(\mathcal{M}) = \text{Tr}_{AdS} (\mathcal{M}) - \text{Tr}_{S}(\mathcal{M})$.
For this specific inner product, the values of $\phi_i$ are 
\begin{equation}
\phi_1 = \phi_5 = c \, , \quad
\phi_2 = 3 a \, , \quad
-\phi_3 = \phi_4 = 6 a \, . 
\end{equation}

\section{The coset action based on $\hat{\mathfrak{g}}$}

The same coset construction of the action for NR strings in AdS given in \cite{Fontanella:2022fjd} may be repeated by taking $\hat{\mathfrak{g}}$ as the Lie algebra of the coset numerator, and $\hat{\mathfrak{h}}$ chosen to be everything in $\hat{\mathfrak{g}}$ except for $\{ H_A, P_a, P_{a'}\}$. The new coset action retains the shape of (\ref{eq:NR_coset_action}), but now any algebraic input information belongs to $\hat{\mathfrak{g}}$. The MC 1-form $A = g^{-1} \dd g$ now requires $g \in \hat{G}$, and the inner product is given in (\ref{inner_product_extended_algebra}). In order to match the general structure of the action of NR strings in string Newton-Cartan backgrounds, one is required to set\footnote{In the matrix representation, this corresponds to $a=0$. The inner product still remains non-degenerate.} $\phi_2=0$ and $\phi_1 = \phi_5$, which leaves the inner product still non-degenerate. The action (\ref{eq:NRAdS}) is then reproduced by the same choice of coordinates as in (\ref{GGK_coset}). In the coordinate-free language, the equations of motion obtained by an arbitrary variation of the group element $g$,  $\delta g = g \xi$ with $\xi \in \hat{\mathfrak{g}}$, can now be derived immediately, without the need of considering component by component as in \cite{Fontanella:2022fjd}, thanks to adjoint invariance of the inner product under the full $\hat{\mathfrak{g}}$. The equations of motion derived in this way are the same as the one found in \cite{Fontanella:2022fjd}, which may be checked by looking at the variation induced on the MC 1-form $\delta A_{\alpha} = \partial_{\alpha} \xi+ [A_{\alpha}, \xi]$, together with the commutation relations of the additional generators $Z_{ab},  Z_{a'b'}$. With the same reasoning, the gauge transformations of the MC 1-form remain the same as before, except that now the MC 1-form has new components in the $Z_{ab},  Z_{a'b'}$ directions, which however are unphysical. As a consequence of this gauge invariance, Noether identities allow us to disregard the equations of motion of unphysical d.o.f. in the same way as in \cite{Fontanella:2022fjd}.   
Finally, the Lax pair argument also remains unaltered.

\section{Conclusions}
In this letter, we showed that Lie algebra expansion naturally suggests that one should consider the non-central extensions $Z_{ab},  Z_{a'b'}$ in the coset action construction of NR strings in AdS. By  extending the algebra, the inner product gains nicer properties, such as adjoint invariance under the full algebra and non-degeneracy. This makes a drastic simplification in the derivation of the equations of motion from the coset action, rendering any further use of the coset action more manageable.   
Moreover, when the generators $Z_{ab},  Z_{a'b'}$ are included, Lie algebra expansion gives the hint on how to construct a finite dimensional representation based on levels, which, on the other hand, cannot be carried out when $Z_{ab},  Z_{a'b'}$ are removed. The gauge fields associated with $Z_{ab},  Z_{a'b'}$ are unphysical, as they play no role in the coset action, leaving the equations of motion, Lax pair and gauge invariance unchanged. The isometry algebra of NR strings in AdS has not been identified yet, and it would be interesting to know if the $Z_{ab},  Z_{a'b'}$ generators are part of it\footnote{The hint comes from the fact the inner product gains nice properties. For NR stings in flat space, where the isometry algebra is known, including $Z_{ab}$ does not make the inner product nicer. However, in that case, it is also easy to show that $Z_{ab}$ cannot be embedded into the global symmetry algebra, as fulfilling the $[G_{Aa}, G_{Bb}]$ relations would imply $[Z_{ab}, J_{cd}]=0$ in the embedding. }.

\section{\label{sec:ackn}Acknowledgments}

We thank A. Torrielli for useful discussions, in particular regarding the level representation. Research at Perimeter Institute is supported in part by the Government of Canada through the Department of Innovation, Science and Economic Development and by the Province of Ontario through the Ministry of Colleges and Universities. AF would like to thank the Department of Physics and Astronomy at the University of Padova and the Department of Mathematics at the University of Surrey for their warm hospitality and for financially supporting his visit during the completion of part of this work. JMNG is supported by the EPSRC-SFI grant EP/S020888/1 \emph{Solving Spins and Strings}.
AF thanks Lia for her permanent support.


\bibliographystyle{nb}

\bibliography{extension_arXiv_v1}

%
%

\end{document}